\magnification = 1200

\def\ref#1{[#1]}
\def\section#1{\bigskip \bigskip {\bf #1} \bigskip}

\centerline{\bf Results and limits of the soliton theory
of DNA transcription}
\footnote{}{{\it Paper presented at the conference {\rm Nonlinear
Phenomena in Biology}, Pushchino, Russia, 22-27 June 1998}}

\bigskip
\centerline{Giuseppe Gaeta}
\medskip
\centerline{\it Dipartimento di Fisica, Universit\'a di Roma}
\centerline{\it 00185 Roma (Italy)}
\centerline{\tt gaeta@roma1.infn.it}

\bigskip\bigskip

{\bf Abstract. } {\it It has been suggested by several
authors that nonlinear excitations, in particular solitary
waves, could play a fundamental functional role in the
process of DNA transcription, providing the opening of
the double chain needed for RNA Polymerase to be able to
copy the genetic code. Some models have been proposed
to model the relevant DNA dynamics in terms of a
reduced number of effective degrees of freedom. Here
I discuss advantages and disadvantages of such an
approach, and discuss in more detail one of the
models, i.e. the one proposed by Yakushevich.} 
\bigskip\bigskip

\section{1. Introduction}

The first step in the replication of DNA \ref{33} is
its transcription, from the original contained in the
cell to a copy -- the RNA messenger -- which will then
be used as a ``master copy'' for producing actual copies
of the genetic information. The evolutionary advantage
of such a messenger is obvious: in this way, the
original DNA is opened -- and thus less protected --
for as small a time as possible.

The transcription process is carried out by a
specialized enzyme, the RNA Polymerase (RNAP); here we
are mostly interested in the dynamical aspect of the
transcription, which is roughly speaking as follows. The
RNAP opens a ''transcription bubble'' of a size of about
15-20 base pairs, and then travels along the DNA chain
keeping the size of the open region more and less
constant, i.e. providing at the same time to open the
chain in front of it and closing back the one behind.

The process undergoes several (even quite long in time)
stops, but in the active phase the RNAP proceeds along
the DNA chain at a speed of 500 -- 1000 base pairs per
second. Since each base pair is linked by two or three
hydrogen bonds, the energy involved in such a process,
even considering only the one to open (and close) the
DNA chain, is of the order of thousands H bonds per
second.

Obviously, as biological systems live at a temperature
of the order of 300 Kelvin, thermal energy is widely
available, but the problem is {\it how is this energy
focused at the right place and with the right timing}
to operate the process.

Needless to say, the problem of energy transport and
focusing is of more general relevance in Biology and
biological systems -- and one indeed which has to be
studied by Biological Physics. Thus, the study of this
problem in the dynamics of DNA transcription is not only
relevant {\it per se} and for the fundamental role of
DNA transcription and then replication in life, but can
also have a more general relevance for the study of
energy transport in living systems.

It has been suggested by several authors that
Nonlinear Dynamics can have a word to say concerning
this, and that energy transport could happen by means
of nonlinear excitations travelling along biological
chains, such as long proteins or indeed DNA. 

The most widely known theory in this context is
probably that of ''Davydov's soliton'' \ref{7,8},
which is of essentially quantum nature. Theories
concerning nonlinear excitations in DNA chains and
their functional role have been proposed by several
authors, following the seminal work of Englander,
Kallenbach, Heeger, Krumhansl and Litwin \ref{11}
(and sometimes it is difficult to understand
priorities, so I will not assign any); these present
differences, but many more common aspects. 

Excellent reviews exist on Nonlinear Waves in
biological systems \ref{34,35}; I would also like to
mention, as a source of a number of inspiring
material on the relations between nonlinear
structures and biology (or more specifically
biological molecules) the volume collecting the
proceedings of the Les Houches 1994 School \ref{28}.

My goal here is not to review and/or compare the
different theories proposed for these would-be
excitations of DNA connected with transcription, but
rather discuss the merit and limitations of the general
idea of what I will call, with a deliberately ambiguous
expression, ''soliton-helped transcription''. 

Roughly speaking, the idea is that there are nonlinear
excitations travelling along the DNA chains, causing a
local opening of the double chain. The RNAP could then
travel along with these, and use the opening of the
chain to read the DNA sequence without having to focus
the energy needed to open the double chain. 

Another advantage of this theory is connected with the
closing of the double chain after the RNAP has passed:
if this is done in a non-coordinated way, it will
generate a substantial quantity of random motion and
thus of thermal energy; if instead this correspond to
the passing of a nonlinear excitation (maybe in this
context it is more appropriate to call our soliton a
``nonlinear coherent structure''), no thermal energy is
generated when the system recovers its local fundamental
state.

I will of course also discuss quantitative results and
predictions, and not just qualitative ideas; in doing
this I will refer to a specific model, the one
suggested by L.V. Yakushevich \ref{39,40} in the late
eighties, on which I have done some work and which is
well suited to analytical investigation. Needless to
say, the very fact that it is amenable to analytic
results means that it is an oversimplified model, and
it is not difficult to find in the literature -- or to
build by oneself -- more detailed models. However, as
I will advocate in more detail in the sequel, I
believe the study of such simple models can teach us a
lot.

Before proceeding, there is a point that should be
made clear for the sake of the more mathematical
reader: the nonlinear excitations we wish to consider
are not necessarily solitons in the proper
mathematical sense \ref{1,9,32}, but rather {\it
solitary waves}; however, it is by now common -- and
more attractive -- to call them {\it solitons}, and I
will conform to this usage. 

Also, the word ``soliton'' is used in two senses, a
topological one \ref{10} together with the dynamical
one \ref{1,9,32}. The solitons to be considered
here will be both solitary waves in dynamical sense
and $S^1$-topological solitons.

\section{2. What should a model contain ?}

The DNA molecule is an awfully (or wonderfully)
complicated one; each segment corresponding to a base
pair has about 100 degrees of freedom, and a DNA
molecule can contain $10^{10}$ (or even $10^{13}$ as
for salamanders) such segments, thus having
$10^{12}$ degrees of freedom. 

As if this was not enough, DNA is very sensitive on the
''boundary conditions'', i.e. to the characteristics
(physical and chemical) of its environment, and no
chemist or biochemist would think of DNA ``in abstract'',
i.e. without specifying in which environment it operates.

With such numbers (even apart from the problem of
modelling the interactions of the DNA molecule with
its environment) it is clear that there is no hope of
detailed mathematical modelling. Actually, these are
so large that we are quite justified in applying the
thermodynamical limit (i.e. consider an infinite
chain), and resorting to statistical mechanics
\ref{22,24}; indeed, several approaches have been
attempted in this direction, and with some success.

However, the dynamical models which have been proposed
to study the dynamics of DNA transcription, only
consider an extremely limited number of degrees of
freedom (one per base in the Yaushevich model); indeed,
it seems hopeless to study a model with more than a very
few degrees of freedom per segment and investigate if it
has solitons, not to say determine the detailed
dynamics of these.

Thus the first question we should ask is: does it make
any sense to study such simplified models of such an
enormously complicated system as DNA ?

Substituting simple mathematical models to complicate real
system is, of course, what has always been done in
theoretical -- and not only theoretical -- Physics, 
and it has been successful many times; but
this is of course not a good reason to say that it can
be done here as well, not to say that in a biological
system the complexity is somehow much more fundamental,
inherent to the system, than in Physics, where we can
e.g. study gravitation without the disturbances of air
by dropping masses in a vacuum tube: in Biology,
``simplifying the system'' most probably means killing it,
and biologists and biochemists are well justified in
their diffidence towards ``Physics-style'' simplified
models.

I would like at this point to open a short parenthesis
to mention the problem of {\it dissociation} (or {\it
melting}) of DNA: experimental data shows there is a 
very sharp transition as temperature is raised, even 
in relatively short specimen of DNA,
leading to dissociation. For
some time this has been a serious obstacle in the
attempt to model DNA as a one-dimensional system,
since one would expect that there is no phase
transition for one-dimensional chains \ref{22,24}. 

However, the theorem on the absence of phase
transitions in one dimension only apply to Ising-like
(spin) systems, while the situation is far more
delicate for systems the state of whose components is
described by a continuous local order parameter.
Basically, in the former case one describes the system
by means of a transfer matrix \ref{22,24} (which is a
Markov one, so relevant restriction on its eigenvalues
exist -- in particular, the fundamental state cannot
be degenerate, and so phase transitions are not
possible), while in the latter one one has to use a
{\it transfer operator} \ref{21}; it is by making use
of this formalism, following work done by J.A. Krumhansl and
J.R. Schrieffer in 1975, that A. Bishop, Th. Dauxois, 
and M. Peyrard proved the existence
of a ''dissociation'' phase transition in DNA
considered as a one dimensional system \ref{4,5};
see also \ref{6,30}.

I do not want to discuss their work here -- among
other reasons, because I could only provide a bad
version of their excellent papers -- but I want to
stress that their success was not only of mathematical
or qualitative nature: their theory compares
successfully to experimental data on the detailed
(spatiotemporal) dynamics of DNA melting; I want to
stress again that they are able to predict not only
average quantities, as it should anyway be the case
with a Statistical Mechanics approach, but a
spatiotemporal pattern: this is a much more severe
test of a model, and the model of Peyrard and Bishop
\ref{29} for DNA melting which was used (with some
minor modifications) in this study did brilliantly
pass the test.

Now, what are the characteristics of this model ?
Well, it is pretty much similar to the models we
consider for transcription, and in particular to the
Yakushevich one. Indeed it models DNA as a
one-dimensional chain, and by singling out
{\it one} degree of freedom per base --
corresponding to ''radial'' displacements along the
axis joining the two bases of a pair -- that is, the
degree of freedom thought to be the most relevant for
the process under study. 

Thus the success of  Bishop, Peyrard and Dauxois in the
study of DNA melting is a witness to the fact that these
-- apparently, hopelessly oversimplified -- models can
indeed provide a good understanding, qualitative but
also quantitative, of the behaviour of DNA in
biologically relevant processes. 
{\it Simple models can be relevant even in very
complicated systems !}

I would take this to the extreme, and say that the
simplest the model, the greater its value: not only
because a simple model can be studied more throughly,
but first and foremost because {\it if} a simple
model works, it means it has focused on the right point,
i.e. that we have reached an understanding of what are
the most relevant aspects (degrees of freedom) of our
system {\it for the process under study}, and thus it
adds more to our knowledge.

Now, conforted by the success of Bishop, Peyrard and
Dauxois, we can look back at the simple counting of
degrees of freedom in DNA with a more optimistic
attitude: it is true that DNA is terribly complicated,
but it is also true that it performs a huge variety of
essential tasks. While the amazement for how seemingly
inconciliable characteristics needed to perform these
different tasks can convive in a single molecule should
remain, it is possible to think of a {\it
process-oriented} modelling of DNA, i.e. to obtain a
model of DNA which does not have any ambition -- or hope
-- to describe it {\it in general}, but only limitedly
to a specific (although relevant) process we want to
study. 

This of course is dependent upon our ability to correctly
identify the specific aspects (or degrees of freedom) of
DNA which are essential for the process we study. This
is the problem whenever we try to formulate a model,
and the Peyrard-Bishop model shows that it is possible
to do this even in DNA.

A less extreme reduction -- keeping a few degrees of
freedom per base rather than only one -- has been
recently considered by Zhang and Collins \ref{46},
again obtaining quite encouraging results. In particular,
they compared results obtained by a simplified 
models with those obtained by an all-atoms simulations,
showing that modelling DNA dynamics in terms of a few
degrees of freedom is by all means satisfactory, at
least for the analysis of specific processes.

Thus, let us come back to discuss what a model of DNA
nonlinear excitations which could play a role in 
transcription should contain.

A first obvious but relevant observation is that the DNA
molecule, like any other one, obeys Quantum Mechanics
(it should be recalled that both the Peyrard-Bishop and
the Yakushevich model are classical); moreover, we are
interested in its behaviour in living condition, i.e. in
particular at a quite precise temperature (around 300
Kelvin, or a bit more for humans). 

Now it happens that at this temperature most of the
degrees of freedom of the atoms constituting the DNA
momlecule will be essentially frozen due to Quantum
Mechanics (their excitation energy being much higher),
and thus many bonds between atoms can be considered as
rigid; on the other side, there are degrees of freedom
whose excitation energy is much lower than the thermal
scale, so that they can be considered to behave
classically.

The consequence of this on modelling is clear: we can
safely consider the degrees of freedom of the first
kind as non-existent, and concentrate on the remaining
-- effective -- degrees of freedom. These should still
be treated quantistically, but for some of them --
those of the second kind -- we can also use a
classical modellization. 

It should be clear that the (quite general) reduction
to ``effective'' degrees of freedom is by no means
sufficient if we want to consider only one -- or even
a few -- degree of freedom per base: we should have
some physical understanding of the process we want to
model in order to understand which are the relevant
degrees of freedom, which cannot always be just a few.

In the case of transcription the relevant process
undergone by the conformation of DNA is its
``unwinding'' \ref{17,28,33} so that the bases, which
usually are stored in the interior of the molecule, as
protected as possible from external agents, come to be
accessible for reading by the RNAP. 

Among the degrees of freedom which are relevant to such
a conformational change, there is one which is
``softer'' than any other, corresponding {\it grosso
modo} to rotations of the bases in a plane orthogonal
to the axis of the double helix. Actually such a
rotation is, obviously, more complicate than this, but
we will investigate the hypothesis that the
comparatively small movements around this main
rotation are not of fundamental importance (see
\ref{17,33}).

It is also reasonable, looking at the geometry of the
molecule, to model the conformational change only in
terms of this; obviously a lot of other degrees of
freedom are involved in such a conformational change,
but the idea is that they will play the role of ``slave
modes'', i.e. there will be small adjustements
involving these but which essentially follow the
relevant changes undergone by the ``master modes'',
those of the fundamental degree of freedom.

Such a way of thinking, in terms of relevant variables
and slave ones, has been successfull in a wealth of
situations, and in Statistical Mechanics
\ref{22,24} or in  Hydrodynamics \ref{25} one has a
very precise way of defining them; however, there is
no {\it a priori} argument showing that this should be
the right way of looking at this specific  problem of
DNA transcription: we can only justify it {\it a
posteriori} if the model works properly.

On the other side, {\it if} the model works, this
means not only that we have the correct values for
coupling constants and so on, but also something more
fundamental: that we have individuated the most relevant
degree of freedom for the process we are studying.

\section{3. The model and its prediction.}

I want now to briefly present the model which has been
proposed by L.V. Yakushevich (with some small
modification), and its predictions concerning the
existence of solitons and their characteristics. 

The discussion of the limitations, or of how acceptable
are a number of approximations it embodies, will be
postponed to the following section.

As I mentioned in the Introduction, several other models
have been proposed; I will mention here \ref{2,23,26,27,
31,37,38,43,44} (and apologize to other
authors I have inadvertentdly forgotten); in many cases 
these give predictions similar to those
of the Yakushevich model, in particular for the
successful ones, but my goal here is not to provide a
review of the different models available, and I prefer
to restrict my discussion to the model I am familiar
with. This is also justified by the fact that this
model has been quite successfull; and I do not think
one can obtain much more without introducing more
detailed models and designing specific experiments to
test their (presumably, only numerical) predictions.
If we look at the Yakushevich ``classification'' of DNA
models \ref{41,42}, I would say that her model provides what is
reasonable to ask to a model of its level, and that
one should now pass to a deeper level to really test
the theory and discriminate among different versions.

In the Yakushevich  model, we consider rotations of the bases
in a plane orthogonal to the double helix axis (which of
course is not completely true), and any other movement
is not considered. Notwithstanding the fact that the
forces involved in hydrogen bonds are highly
directional, the forces among the bases are considered
to be linear in the distance $d$ between the extremities
interacting to form the H bonds. This will produce a
nonlinear dependence on the rotation angles.

An even more severe approximation is introduced
concerning the characteristics of the bases and their
interactions: all bases are considered as identical, and
so the same also holds for base-pair interactions. 

Let us consider the interaction among bases on the two
chains at site $n$. If $r$ is the distance from the center
of rotation to the site responsible for the intra-pair
interaction, $\theta_1 (n)$ and $\theta_2 (n)$ the angles
(both measured in, say, counterclockwise direction
\ref{14}) of rotation, and $L$ the distance at rest,
we would then have for the distance $d$,
$$ d^2 \ = \ [ 2L + 2 r - r (\cos \theta_1 + \cos
\theta_2 ) ]^2 + [ r (\sin \theta_1 + \sin \theta_2 )
]^2 \ . \eqno(1) $$
However, in the Yakushevich model one makes the approximation
$L=0$ (see next section).

Thus the potential ``rotation'' energy will be $V_r =
(K_r/2) d^2$, with $K_r$ a real constant and $d^2$ the
simplified expression corresponding to taking $L=0$ in
(1). Notice that the force this produces is nonlinear in
$\theta_1 , \theta_2$.

Together with this, we will have an elastic interaction
with nearest neighbour bases along the two chains, with
interaction ``torsional'' energy $V_t = (K_t /2) (\theta_1
(n) - \theta_1 (n-1) )^2$ and the like for the other
chain and for the interaction with bases at site $(n+1)$.

There is another kind of interaction which should be
taken into account\ref{3,12}: bases which are
half-pitch of the helix apart are actually near enough
in ambient space, and they interact through the
formation of water filament; the energy associated to
this ``helicoidal'' interaction is $V_h = (K_h /2 ) [
\theta_1 (n+p) - \theta_2 (n) ]^2$ and the like for
the other chain, interchaninging indices 1 and 2, and
for interaction with bases at site $(n-p)$. Here $p$
is the half-pitch of the helix measured in units of
base sites, which for DNA means $p=5$. This
interaction, although weaker than the other ones,
produces a qualitative change in the dispersion
relations and in the predictions of the model.

Thus, introducing also the kinetic energy 
$K = I/2 {\dot \theta}_1 (n) + I/2 {\dot \theta}_2
(n)$, we can finally write the total energy for the
(Yakushevich model of the) DNA double chain:
$$ \eqalign{
H \ = &\  \sum_n \, {1 \over 2} I \, \left[ {\dot \theta}_1^2 (n) +
{\dot \theta}_2^2 (n) \right] \ + \cr
 & + \ \sum_n \, {1 \over 2} K_t \,
\left[ \left( {\theta}_1 (n) - {\theta}_1 (n-1) \right)^2 \, + \,
\left( {\theta}_2 (n) - {\theta}_2 (n-1) \right)^2 \right] \ + \cr
 & + \ \sum_n \, {1 \over 2} K_r \,
\left[Ê6 r^2 + 2 r^2 \cos (\theta_1 (n) - \theta_2 (n) )
- 4 r^2 ( \cos (\theta_1 (n) ) + \cos (\theta_2 (n) )
) \right] \ + \cr
 & + \ \sum_n \, {1 \over 2} K_t \,
\left[Ê\left( {\theta}_1 (n) - {\theta}_2 (n-p)
\right)^2 \, + \,  \left( {\theta}_2 (n) - {\theta}_1
(n- p) \right)^2 \right]  
 \ . \cr} \eqno(2) $$

>From this we can easily derive the equations of
motion for the system; these are more usefully written
in terms of the variables
$$ \psi_n := {{\theta}_1 (n) + {\theta}_2 (n) \over
2} \ , \ \chi := {{\theta}_1 (n) - {\theta}_2 (n)
\over 2} \eqno(3) $$ 
and are 
$$ \eqalign{
{\ddot \psi}_n \ = &\ - \alpha \sin \psi_n  \cos
\chi_n  \, + \, \beta \left( \psi_{n+1} - 2 \psi_n +
\psi_{n-1} \right) \, + \, \gamma \left( \psi_{n+p} - 2 \psi_n +
\psi_{n-p} \right) \cr 
{\ddot \chi}_n \ = &\ - \alpha \sin \chi_n (
\cos \psi_n - \cos \chi_n ) \, + \, \beta \left(
\chi_{n+1} - 2 \chi_n + \chi_{n-1} \right) \, - \, \gamma
\left( \chi_{n+p} + 2 \chi_n + \chi_{n-p} \right) \cr}
 \eqno(4) $$
(notice the sign differences in the $\gamma$ terms),
where $\alpha = K_r / I $, $\beta = K_t /I$ and
$\gamma = K_h/I$. 

In these equations we find the different elastic
constants $K_r , K_t , K_h$ or equivalently $\alpha
, \beta , \gamma$; these could be seen as free
parameters to be fitted, but it is also possible to
give an estimate on their values based on first
principles, i.e. on an estimate of the strength of the
interactions they should model. The values given in
\ref{14,17} are as follows: 
$$ K_r = 0.13 \, {\rm eV/rad^2} \ , \ K_t = 0.025 \, {\rm
eV/rad^2} \ , \ K_h = 0.009 \, {\rm eV/rad^2} \ , \ I = 3
\cdot 10^{-37} \, {\rm cm}^2 {\rm g} \ . \eqno(5) $$

Now, in the analysis of the Yakushevich model it has been usual
to pass to the continuum approximation, i.e. having
fields $\psi (x,t) , \chi (x,t)$ instead of the
sequences of variables $\{ \psi_n (t) \}$ and $\{
\chi_n (t) \}$; in this way the chains of ODEs (4) are
replaced by coupled PDEs 
$$ \eqalign{
\psi_{tt} \ = & \ - \alpha \sin \psi \cos \chi +
\beta \delta^2 \psi_{xx} + \gamma {\cal W}_+
(\psi ) \cr
\chi_{tt} \ = & \ - \alpha \sin \chi (\cos \psi -
\cos \chi ) + \beta \delta^2 \chi_{xx} - \gamma
{\cal W}_- (\chi ) \cr} \ , \eqno(6) $$
where $\delta$ represents the distance between
consecutive bases along the double chain axis
($\delta \simeq 3.4$ Angstroms), and the nonlinear
terms ${\cal W}_\pm$ are defined by
$$ {\cal W}_\pm (\phi ) \ := \ \phi (x + p \delta ,
t) \mp 2 \phi (x,t) + \phi (x- p \delta , t) \ .
\eqno(7) $$ 

This passage to the continuum is questionable, and
indeed in their study of the melting transition
Peyrard and Dauxois have pointed out interesting
effects due precisely to the discrete nature of the
model (there is a {\it Peierls barrer} restraining
energy flow); we will discuss this point in next
section as well.

We can then analyze (6); first of all -- considering
the linear terms -- we obtain the dispersion relations
for the $\psi$ and the $\chi$ branch; these are
respectively 
$$ \eqalign{ \omega_\psi^2 \ = & \ \alpha + \beta
\delta^2 q^2 + 4 \gamma \sin^2 (p \delta q /2 ) \cr
\omega_\chi^2 \ = & \ \beta \delta^2 q^2 + 4 \gamma
\cos^2 (p \delta q /2 ) \ . \cr} \eqno(8) $$

Notice that now (due indeed to the introduction of
the ``helicoidal'' interactions) the minimum of the
dispersion curve, and thus the threshold for
excitation of linear modes, corresponds in terms of
the natural wavenumber unit $\xi = p \delta q/2$, to
$\xi = 1.4 $; the corresponding wavelength will be the
first to be excited and should act as a seed for the
formation of nonlinear structure. Thus, it gives a
rough estimate for the size of excitations, which
is $\lambda \simeq 11 \delta$. 

Similarly, the corresponding $\omega = \omega (\xi_0
)$ gives a rough estimate for the timescale of
small oscillations, which is of 57 ps, and
similarly one obtains an estimate of the amplitude
of ``linear'' oscillations, which is $0.77 \pi$. This
seems sufficient to ignite the unwinding and the
nonlinear regime.

We stress that the minima of the dispersion relations
are essentially dependent on the helicoidal
interaction term (without this we actually would have
phonons in the model). Data for linearized dynamics
are in a way even more significant when we analyze
DNA melting rather than transcription, but this lies
outside our scope here; see \ref{14,17} for a
discussion. Also, the presence of helicoidal
interactions lead to crossings between the two
branches of the dispersion relations, which could play
a role in energy transfers \ref{15}.

It is remarkable that the estimate of the size of
collective motions is of the same order of magnitude
as experimental data on the size of transcription
bubbles. Again, for melting models the agreement is
significant also for other relevant quantities,
again providing the right order of magnitude.

Let us now look more precisely to solitons, or
solitary waves, in (6). These will be solution of
the form $\phi (x,t) = \phi (x - vt) \equiv \phi
(z)$, where $\phi = \psi , \chi$. Moreover they
should satisfy 
$$ \lim_{z \to \pm \infty} \phi_z \, = \, 0 \ , \ 
\lim_{z \to \pm \infty} \phi \, = \, 2 m_\pm \pi
\eqno(9) $$ in order to have a finite energy. They
can thus be characterized in terms of their winding
numbers $\mu = (m_+ - m_-)$ for $\psi$ and $\chi$
\ref{13,14,17}; these corresponds to the phase shift
of the solitons.

The lowest energy solitons will be those with 
$\mu = (1,0)$ and $\mu = (0,1)$, and we concentrate
on these. 

If we set the helicoidal interaction to zero, it is
possible to determine exact solitary wave solutions.
These are given by
$$ \psi (z) \ = \ {\rm arccos} \, \left( {\sinh^2 (a z) -
1 \over \sinh^2 (a z) + 1 } \right) \eqno(10) $$
and $\chi (z) = 0$ in the case of the $(1,0)$
soliton, while the $(0,1)$ soliton is given by
$\psi (z) = 0$ and
$$ \chi (z) \ = \ {\rm arccos} \, \left( { a^2 z^2 - 1
\over a^2 z^2 + 1 } \right) \ \ . \eqno(11) $$
Here $a = [2 \alpha / (\beta \delta^2 - v)]^{1/2}$,
where $v$ is the soliton's speed.

A numerical integration \ref{13} shows that the
introduction of the (weak) helicoidal terms does not
really alter the fully nonlinear structures.

The solitonic solution can exist, in principle, with
any speed $v < \beta \delta^2$. 
Moreover, the dependence of their
energy on the speed \ref{13,17} is given by a
relation of the type $ E = c_1 / (c_2 - v^2 )$, as it
should for waves, so that up to high speed the energy
is very little sensitive to speed; see \ref{13} for
details. This means that in this model there is no
selection of soliton speed. A refinement of the model
\ref{16} seems to be able to provide such a selection
based on some  conditions at interface of A-T and G-C
regions, but in this model the soliton can not
actually exist in proper terms, i.e. as solitary waves
travelling along the chain with no dispersion.

As for the size of the soliton, this will correspond to 
the size $s$ of the region in which $\psi$ differs from its 
asymptotic value by more than a given quantity $\varepsilon$,
and thus will depend on this. We call $z_+ , z_-$ the values
of $z$ at which $\psi (z_- ) =  \varepsilon$, $\psi(z_+) =
2 \pi - \varepsilon$. For $a z >> 1$, the argument of the 
arccos in (10) is $f(z) \simeq 1 - 2 \exp [- 2 a z]$, and thus 
if $\cos  (\varepsilon  ) = 1  - \lambda (\varepsilon )$, we
have to solve $\lambda (\varepsilon ) = 2 \exp [- 2 a z_\pm ]$;
obviously $z_- = - z_+$, and so $s =  | a^{-1} \ln [\lambda 
(\varepsilon ) / 2] |$.

As for $a$, when $v=0$ we get $a = \sqrt{2 \alpha / ( \beta \delta^2 )} 
:= a_0 \delta^{-1}$, and the soliton size measured in base sites units
is thus
$$ s \ = \ \left\vert {1 \over a_0 } \, \ln [ \lambda (\varepsilon ) / 2 ]
\right\vert \, \delta \ . \eqno(12) $$
If we choose $\varepsilon = 5^o \simeq 0.044$, we get $\lambda 
(\varepsilon ) \simeq 2.9 \cdot 10^{-7}$ and 
$\ln [ \lambda (\varepsilon ) / 2 ] \simeq 15.75$; with the values 
given in (5) we get $a_0 \simeq 3.22$ and thus it results $s \simeq 5 
\delta$.

This is smaller than the observed size of the transcription bubble, 
which is of order  15 - 20 bases, but again we have got the right 
order of magnitude.

\section{4. Limitations of the model.}

In the previous section we have seen that the
Yakushevich model, with the introduction of
``helicoidal'' terms, is able to give a prediction of
the order of magnitude of several physical quantities
for coherent structures and excitations in the DNA
chain. This is specially significative when we
consider the crude approximations which have been
introduced in the formulation of the model, so that we
are encouraged to think that the model does indeed
correctly identify the essential degrees of freedom
for the dynamics related to the transcription process.

In this section we want instead to emphasize the
limitations and the shortcomings of the model; in
most cases this will actually be not specific to the
Yakushevich model, but common to the whole set of
simple models we have advocated above.

We will actually first of all mention a shortcoming
of teh Yakushevich model itself, which was noticed
by Gonzalez and Martin Landrove \ref{18}: the 
seemingly unoffensive approximation $L=0$, see
eq.(1), is actually selecting a special case: any
nonzero value of $L$ would cause qualitative (and
not only quantitative) differences in some basic
behaviour of the model; in particular the pairing interaction
would be through nonlinear terms only, with no linear ones. As
this situation is unphysical, we see that the apparently special 
choice $L=0$ does actually produce a ``generic'' behaviour; this
has also suggested to consider the Yakushevich model as an 
effective one.

Having mentioned this fact, we can pass to aspects
which are not really specific of the Yakushevich model,
although we will refer to it for the sake of definiteness.

A first obvious limitation is in the fact that bases
are considered as identical. Needless to say, this
is not the case. If one could be tempted to think
that distinguishing between Purines and Pyrimidines
only could be acceptable, as a first approximation,
in view of the modellization adopted (the physical
quantities entering in the model are homogeneous
enough for the two purines and for the two
pyrimidines, see \ref{16}), considering all bases
as equivalent is highly questionable: e.g., masses
range from 110 to 150 atomic mass units, and inertia
moment from 1500 to 2500 m.u. Angstrom. 

Also, the energies involved in the base-base
interactions are quite different for A-T and C-G
pairs: the H-bond pairing energy is of 7.0 kcal/mole
in the first case and 16.8 kcal/mole in the second
one; the opening energy is respectively of 4.0 and
7.5 kcal/mole, and the stacking energies in the case
of a homogeneous sequence is of 5.4 kcal/mole for a
sequence of A-T (or T-A) pairs, and of 8.3 kcal/mole
for G-C (or C-G) pairs. 

This calls immediately for a more realistic
modellization of the DNA chain, taking into account
the different characteristics of different bases. It
is quite clear, however, that renouncing to the
spatial homogeneities of the chain (and to any kind
of spatial periodicity as well if we want to model
realistic sequences) means that such more realistic
models will not have proper solitary waves, i.e.
solutions of the form $\phi (x,t) = \phi (x - vt)$
or the discrete analogue.

Naturally, from the point of view of biological
significance of solutions we do not need that these
are exactly solitary waves propagating with no
dispersion and/or change in shape: it is sufficient
that they are sufficiently stable over sufficiently
long space and time intervals. 

Obviously an analytical investigation of a realistic
model with a true base sequence is outside reach, so
this should be studied numerically. As far as I know,
such a numerical study has never been conducted --
opposite to the detailed studies conducted by Dauxois
and Peyrard for the related modelling of the
dissociation process by a realistic model based on
the Peyrard-Bishop one -- and so we cannot draw any
conclusion in this respect (I do not know of any
apriori obstacle to such a numerical study; in
particular, it seems that for the nonlinear
structure the weak helicoidal interaction, which was
essential at the level of dispersion relations, has
very little influence, so that one could study a
simple ``planar'' model \ref{17}).

Partial informations can be extracted for special
kind of base sequences (at least in the
purine-pyrimidine approximation); thus, for
example, alternating regions such as ``TATA boxes''
would select a zero soliton speed \ref{16}. This is
an interesting result, as TATA boxes are indeed know
to code the pausing sites for the transcription
process.

However, for more general sequences, one has reasons
to think that a more detailed Yakushevich-like model
could actually be {\it worse} than the original
simple one from the point of view of correspondence
to experimental observations; this could be due to
the fact that the structures we are interested in
involve several base pairs, so that there could be a
mechanism of ``effective self-averaging'' in the
dynamics \ref{16}.

Another relevant limitation of Yakushevich-like
models is that there seem to be no selection of the
soliton speed (see however below): solitons can
exist at all speed below the maximal one $v_{\rm max} 
= \beta \delta^2$, and energy considerations will
not lead to a real selection among low speeds. This
is not really encouraging, as transcription speed is
not enormously fluctuating: we would expect there is
some mechanism to regulate the soliton speed. This
would also be essential to the functional role which
the soliton is supposed to have, as it is necessary
in this picture that the soliton speed is compatible
with the speed of the transcription by the RNAP.

One should also recall that, although most of
the investigation of the Yakushevich model have been
based on the continuum approximation, such an
approximation is not justified: actually, as
mentioned above, in the case of DNA denaturation
Dauxois and Peyrard have pointed out the essential
role played by discreteness of the chain. 

Again, it seems that no thorough investigation of
the Yakushevich model on a chain (even an
homogeneous one) have been conducted, so that
properly speaking we cannot be sure about the
existence and characteristic of soliton-like
excitations in this case.

Another relevant point to be recalled is that a
soliton in a continuous homogeneous medium would move
with zero energy barrer. Essentially, this can be seen
realizing that the field Lagrangian will involve
$\phi_t^2$ and $\phi_x^2$, and when $\phi = \phi (x
- vt)$ these two terms become $\phi_z^2$ and
$v^2 \phi_z^2$: thus a soliton field
configuration with zero speed will have an energy
$E_0$, and one with a speed $v = \varepsilon$, an
energy $E' = (1 + \varepsilon^2 ) E_0$.

However, the discrete structure of the chain lead to an
effective barrer (known as Peierls barrer) for the
motion. The result is that excitations of sufficiently
low energy can be pinned on-site precisely by the
presence of this barrer and thus by the discrete
structure. This suggests that in a discrete
Yakushevich model only solitons of sufficient high
energy and thus speed could actually move along the
chain, passing the Peierls barrer. 

It should be noticed that this could provide the
required mechanism for soliton speed, imposing $v >
v_0$; the significance of this would of course
depend on the actual value of $v_0$ and how it
compares with the observed speed of the 
transcription bubble.

Anotrher major limitation of the Yakushevich -- and
similar -- models is precisely the fact of focusing on 
the rotational motion connected with the reading of 
the DNA sequence by RNAP: indeed, the rigidity (or
softness) of the concerned degree of freedom is quite
comparable to the one of ``radial'' motion as considered
in the Peyrard-Bishop model; thus, it would be highly 
desirable to consider these two degrees of freedom together,
and combine the Peyrard-Bishop and the Yakushevich models
into a single one.

Finally, the models -- such as Yakushevich's
one -- studied in connection with the problem of DNA
transcription are all purely deterministic and only
concern the DNA chain itself, i.e. with no
interaction with its environment. One hopes that in
a large part the chemico-physical parameters
describing the environment could somehow be embodied
in effective values of the constants defining the
model (or one could limit to consider ``standard''
environment), but however the DNA has to work at
finite, actually quite different from zero,
temperature: thus unavoidably one has to consider
thermal energy of the environment and energy
exchanges involving this. 

In considering this problem, one could resort to
Statistical Mechanics considerations (which lie
outside the scope of the present short report); or
more in the spirit of Yakushevich-like models one
should introduce random forces modelling thermal
effects. It is not clear, nor -- as far as I am
aware -- it has been studied if the solitons would
survive the introduction of such random forces. 

On the other side, the role of thermal energy is not
necessarily negative from the point of view of
existence of solitons: it is well possible that these
same random forces would help the initial formation of
nonlinear sructures by pumping energy into the more
easily excitable modes until they ignite the formation
of soliton-like collective structures.

It is maybe worth mentioning that statistical
approaches to inhomogeneous models have been also
considered recently; I will just mention the work
of Hisakado and Wadati \ref{19}, where the distribution
of masses and physical constants of the bases is a
gaussian white noise, and the more specific work
of Homma \ref{20}, which sets the basis for a 
statistical mechanics treatment of double chain
models with sine-like nonlinearities, i.e. of the
kind we consider here.

In this respect, it should be recalled that the
whole process of transcription is strongly dependent on
temperature (as is also that of denaturation), and in
{\it in vitro} experiments where temperature is varied
the transcription bubble forms sharply in a very small
temperature range, see \ref{17} and references
therein: thus we are compelled to introduce
dependence upon temperature in any modellization of
the process. 

This problem is considered e.g. in \ref{36}; see also
\ref{45} for consideration of the environment in the 
somehow related problem of a single polyethilene chain.

\section{5. Summary and conclusions.}

I have given a short discussion of the idea that
nonlinear excitations could play a role in the
process of DNA transcription, i.e. that the
transcription bubble could correspond to a solitary
wave travelling along the chain, which the RNAP
could then ``surf'' in order to access the base
sequence with no energy to provide for opening the
double helix. 

I have discussed at some length the general idea of
providing a simple model for a specific DNA process,
and argued that despite the tremendous complexity of
the DNA model this approach is not bound to fail.
I have then recalled the main features of the model
proposed by Yakushevich, mentioning some encouraging
achievements and several limitations. 

These limitations, however, more than being inherent
to the model are limitation of the studies conducted
so far. It is clear that the model is too simple to be
valid as it is, and that it is needed to go ``one step 
further'' in the Yakushevich classification of DNA
models \ref{41,42}, but only a more thorough analysis can
focus on the detailed refinements which are needed.

In particular, I have pointed out several directions
in which I believe it is necessary to generalize the
model and to investigate its behaviour, such as 
considering real base sequences and
thermal effects.

I hope this conference can, among other things, also
stimulate a reflection on this theme, and maybe new
works to analyze more realistic models of the
nonlinear dynamics involved in DNA transcription.

\section{References.}

\parindent=30pt
\def\bitem#1#2{\item{#1.} }

\bitem{1}{Cal} F. Calogero and A. Degasperis, {\it
Spectral transforms and solitons}, North Holland
(Amsterdam), 1982

\bitem{2}{CM} P.L. Christiansen and V. Muto, {\it
Physica D} {\bf 83} (1993), 93

\bitem{3}{Dau} Th. Dauxois, {\it Phys. Lett. A} {\bf
159} (1991), 390

\bitem{4}{DP} Th. Dauxois and M. Peyrard, {\it Phys.
Rev. Lett.} {\bf 70} (1993), 3935

\bitem{5}{DPB} Th. Dauxois, M. Peyrard and A.R. Bishop,
{\it Phys. Rev. E} {\bf 47} (1993), R44 and 684

\bitem{6}{DPW} Th. Dauxois, M. Peyrard and C.R. Willis,
{\it Physica D} {\bf 57} (1992), 267

\bitem{7}{Da1} A.S. Davydov, {\it J. Theor. Biol.} {\bf
38} (1973), 559

\bitem{8}{Da2} A.S. Davydov, {\it Biology and Quantum
Mechanics}, Pergamon (London), 1982

\bitem{9}{Dra} P.G. Drazin and R.S. Johnson, {\it
Solitons: an introduction}, Cambridge University
Press (Cambridge), 1989, 1993

\bitem{10}{DNF} A. Dubrovin, S.P. Novikov and A.
Fomenko, {\it Modern Geometry}, Springer (Berlin),
1984

\bitem{11}{Eng} S.W. Englander, N.R. Kallenbach, A.J.
Heeger, J.A. Krumhansl and S. Litwin, {\it Proc.
Natl. Acad. Sci. U.S.A.} {\bf 777} (1980), 7222

\bitem{12}{Ga1} G. Gaeta, {\it Phys. Lett. A} {\bf 143}
(1990), 227

\bitem{13}{Ga2} G. Gaeta, {\it Phys. Lett. A} {\bf 168}
(1992), 383

\bitem{14}{Ga3} G. Gaeta, {\it Phys. Lett. A} {\bf 172}
(1993), 365

\bitem{15}{Ga4} G. Gaeta, {\it Phys. Lett. A} {\bf 179}
(1993), 167

\bitem{16}{Ga5} G. Gaeta, {\it Phys. Lett. A} {\bf 190}
(1994), 301

\bitem{17}{RNC} G. Gaeta, C. Reiss, M. Peyrard and Th.
Dauxois, {\it Rivista del Nuovo Cimento} {\bf 17}
(1994), n.4

\bitem{18}{GML} J.A. Gonzalez and M. Martin Landrove,
{\it Phys. Lett. A} {\bf 191} (1994), 409

\bitem{19}{HWa} M. Hisakado and M. Wadati, {\it J. 
Phys. Soc. Jap.} {\bf 64} (1995), 1098

\bitem{20}{Hom} S. Homma, {\it Physica D} {\bf 113} 
(1998), 202

\bitem{21}{KS} J.A. Krumhansl and J.R. Schrieffer, {\it
Phys. Rev. B} {\bf 11} (1975), 3535

\bitem{22}{LL} L.D. Landau and E.M. Lifsitz, {\it
Statistical Physics}, Pergamon (London), 1955

\bitem{23}{LF} V. Lisy and V.K. Fedyanin, {\it J. Biol.
Phys.} {\bf 18} (1991), 127

\bitem{24}{Ma} S.K. Ma, {\it Statistical Mechanics},
World Scientific (Singapore), 1985

\bitem{25}{Man} P. Manneville, {\it Dissipative
structures and weak turbulence}, Academic Press
(N.Y.), 1990

\bitem{26}{MLC} V. Muto, P.S. Lomdahl and P.L.
Christiansen, {\it Phys. Lett. A} {\bf 136} (1989),
33; {\it Phys. Rev. A} {\bf 42} (1990), 7452

\bitem{27}{MSC} V. Muto, A.C. Scott and P.L.
Christiansen, {\it Physica D} {\bf 44} (1990), 75

\bitem{28}{Pey} M. Peyrard ed., {\it Nonlinear excitations
in biomolecules}, Springer (Berlin), 1995

\bitem{29}{PB} M. Peyrard and A.R. Bishop, {\it Phys.
Rev. Lett.} {\bf 62} (1989), 2755

\bitem{30}{PDH} M. Peyrard, Th. Dauxois, H. Hoyet and
C.R. Willis, {\it Physica D} {\bf 68} (1993), 104

\bitem{31}{Pro} E.W. Prohofsky, {\it Phys. Rev. A} {\bf
38} (1988), 1538 and 5332

\bitem{32}{Rem} M. Remoissenet, {\it Waves called
solitons}, Springer (Berlin), 1994

\bitem{33}{Sae} W. Saenger, {\it Principles of nucleid
acid structure}, Springer (Berlin) 1984, 1988

\bitem{34}{Sc1} A.C. Scott, {\it Phys. Rep.} {\bf
217} (1992), 1

\bitem{35}{Sc2} A.C. Scott, {\it Solitary waves in
Biology}, in \ref{28}

\bitem{36}{Sta} E.B. Starikov, {\it Mol. Biol.} {\bf 24}
(1990), 1194

\bitem{37}{TH} S. Takeno and S. Homma, {\it Progr.
Theor. Phys.} {\bf 70} (1983), 308; {\bf 72} (1984),
679; {\bf 77} (1987), 548

\bitem{38}{TDP} M. Techera, L.L. Daemen and E.W.
Prohofsky, {\it Phys. Rev. A} {\bf 40} (1989),
6636; {\bf 42} (1990), 1008 

\bitem{39}{Ya1} L.V. Yakushevich, {\it Studia Biophys.}
{\bf 121} (1987), 201

\bitem{40}{Ya2} L.V. Yakushevich, {\it Phys. Lett.}
{\bf 136} (1989), 413

\bitem{41}{Ya3} L.V. Yakushevich, {\it Quart. Rev.
Biophys.}  {\bf 26} (1993), 201

\bitem{42}{Ya4} L.V. Yakushevich, {\it Physica D} 
{\bf 79} (1994), 77

\bitem{43}{Yom} S. Yomosa, {\it J. Phys. Soc. Jap.} {\bf
51} (1982), 3318; {\it Phys. Rev. A} {\bf
27} (1983), 2120; {\bf 30} (1984), 474

\bitem{44}{Zan} L.L. van Zandt, {\it Phys. Rev. A} {\bf
40} (1989), 6134; {\bf 42} (1990), 5036

\bitem{45}{ZC1} F. Zhang and M.A. Collins, {\it Phys. Rev. E}
{\bf 49} (1994), 5804

\bitem{46}{ZC2} F. Zhang and M.A. Collins, {\it Phys. Rev. E}
{\bf 52} (1995), 4217

 \bye